\documentclass{iopart}

\usepackage{graphicx}
\usepackage{amssymb}
\usepackage{ifthen}
\usepackage{hyperref} 
\usepackage{color}

\definecolor{darkgreen}{rgb}{0,.7,0}
\definecolor{linkblue}{rgb}{0.,0.,0.9333}

\newcommand{\TITLE}{Quenching and Tomography from RHIC to LHC}

\hypersetup{
    pdffitwindow=true,      
    pdftitle={\TITLE},    
    pdfauthor={W. A. Horowitz and Miklos Gyulassy},     
    pdfnewwindow=true,      
    bookmarksnumbered=true,
    colorlinks=true,       
    linkcolor=red,          
    citecolor=darkgreen,        
    filecolor=magenta,      
    urlcolor=linkblue,           
    menucolor=blue,
}

\begin{document}

\title{\TITLE}
\author{W.~A.~Horowitz}
\address{Department of Physics, University of Cape Town, Private Bag X3, Rondebosch 7701, South Africa}
\ead{wa.horowitz@uct.ac.za}
\author{Miklos Gyulassy}
\address{Department of Physics, Columbia University, 538 West 120th Street, New York, NY 10027, USA}
\ead{gyulassy@phys.columbia.edu}

\begin{abstract}
We compare fully perturbative and fully nonperturbative pictures of high-$p_T$ energy loss calculations to the first results from LHC.  While oversuppressed compared to published ALICE data, parameter-free pQCD predictions based on the WHDG energy loss model constrained to RHIC data simultaneously describe well the preliminary CMS hadron suppression, ATLAS charged hadron $v_2$, and ALICE $D$ meson suppression; we also provide for future reference WHDG predictions for $B$ meson $R_{AA}$.  However, energy loss calculations based on AdS/CFT also qualitatively describe well the RHIC pion and non-photonic electron suppression and LHC charged hadron suppression.  We propose the double ratio of charm to bottom quark $R_{AA}$ will qualitatively distinguish between these two energy loss pictures.
\end{abstract}

\emph{Introduction} The current and future challenge to theoretical calculations will be to simultaneously describe the multiple high-$p_T$ observables measured at RHIC and LHC \cite{Bathe,Masui:2011qi,Schukraft:2011cz,Wyslouch,Steinberg}.  This will prove to be a daunting task as a simultaneous description of multiple observables at RHIC is not even at hand \cite{Jia:2011pi}.  Furthermore, generic arguments based on RHIC results naturally lead one to predict a suppression greater than that in published LHC data, regardless of $L$, $L^2$, or $L^3$ pathlength (i.e.\ specific energy loss mechanism) dependence \cite{Horowitz:2011gd}.  Despite the difficulty of the task, the reward is potentially enormous: high-$p_T$ particles provide the most direct probe of the $Q^2$ dependence of the organization of the soft-soft and soft-hard degrees of freedom in the quark-gluon plasma (QGP) created in heavy ion collisions.  In this work we will compare the fully weakly- and strongly-coupled pictures to the recent data from RHIC and LHC; we will find that while neither provides a fully compelling picture of the experimental results, neither is ruled out by data.

\begin{figure}[!htb]
\centering
$\begin{array}{cc}
\includegraphics[width=2.2in]{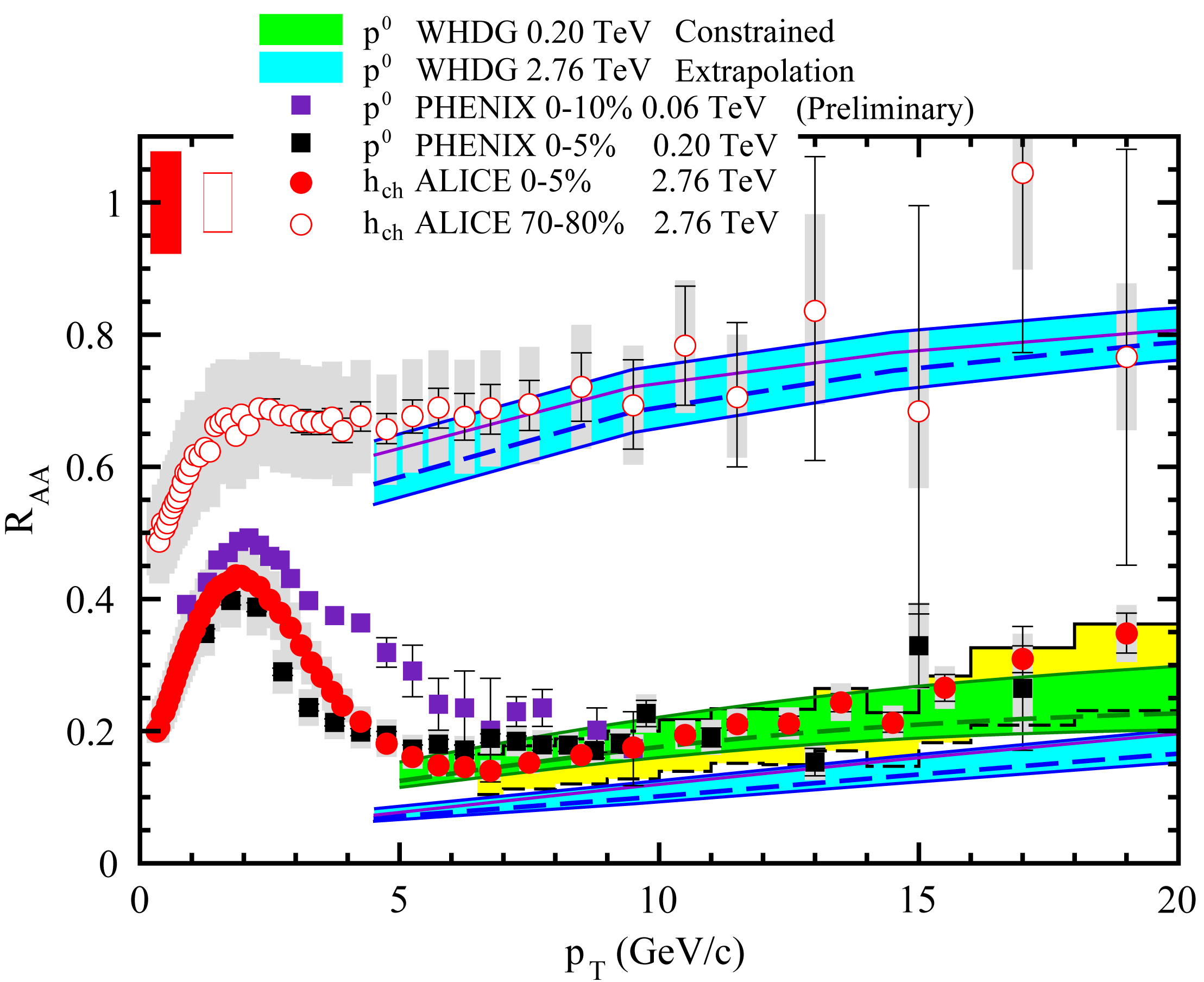} &
\includegraphics[width=1.8in]{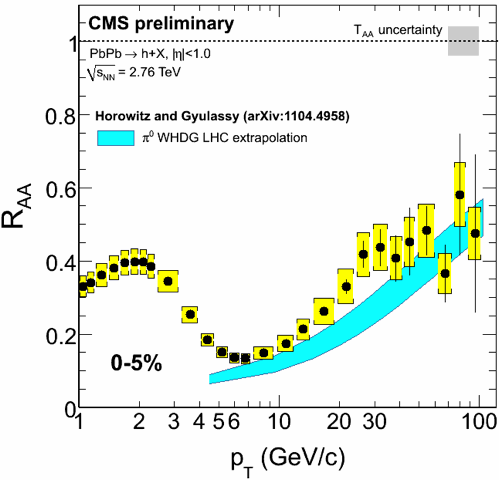}
\end{array}$\\
\begin{flushleft}
\vspace{-.2in}
$\begin{array}{ll}
\mbox{\hspace{.5in}\footnotesize(a)} & \mbox{\hspace{2.2in}\footnotesize(b)}
\end{array}$
\end{flushleft}
\vspace{-.1in}
\caption{\label{RAA}(a) $R_{AA}(p_T)$ for $\pi^0$ at $\surd s$ = 62.4 and 200 AGeV at RHIC \cite{Purschke} and $h_{ch}$ at $\surd s$ = 2.76 ATeV at LHC \cite{Aamodt:2010jd}. A rigorous extraction of $dN_g/dy$ was performed on the $\surd s$ = 200 AGeV RHIC data for the WHDG model \cite{Adare:2008cg}: the best-fit curve is given by the dashed green line and the one standard deviation uncertainty by the green band; these results were extrapolated to LHC energies, with the best fit curve given by the dashed blue line and the one standard deviation uncertainty by the blue band.  In (b) this same blue band is compared to preliminary CMS data \cite{CMSRAA}.}
\end{figure}

\begin{figure}[!htb]
\centering
$\begin{array}{cc}
\includegraphics[width=1.8in]{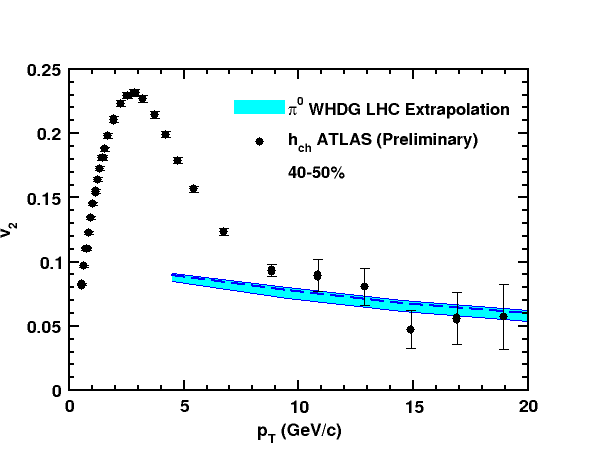} &
\includegraphics[width=2.5in]{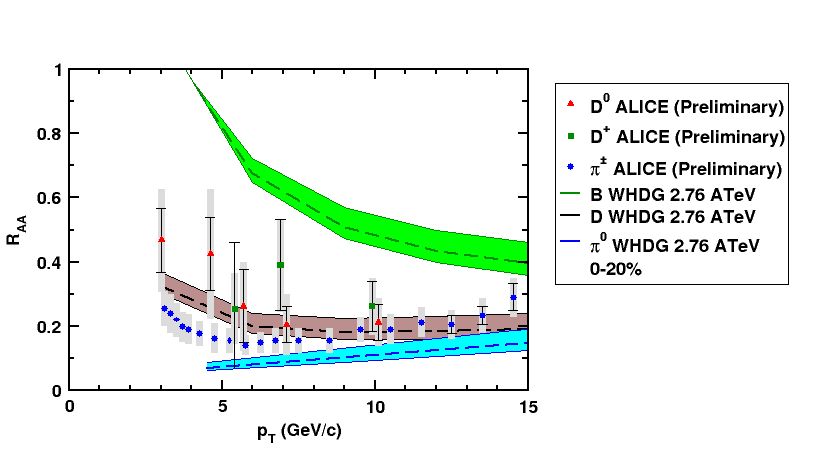}
\end{array}$\\
\begin{flushleft}
\vspace{-.2in}
$\begin{array}{ll}
\mbox{\hspace{.5in}\footnotesize(a)} & \mbox{\hspace{2.2in}\footnotesize(b)}
\end{array}$
\end{flushleft}
\vspace{-.1in}
\caption{\label{v2RAA}(a) Comparison of ATLAS charged hadron $v_2(p_T)$ \cite{Jia} to WHDG predictions at 40-50\% centrality. (b) Comparison of ALICE $D^0$, $D^+$, and $\pi^\pm$ mesons \cite{Dainese:2011vb} to WHDG predictions at 0-20\% centrality. 
}
\end{figure}

\emph{Perturbative Picture} In a perturbative calculation of partonic energy loss in a QGP, the high-$p_T$ parent parton loses energy via both elastic and inelastic processes; the size of this loss in calculable using standard Feynman diagram techniques. This in-medium energy loss is convolved with the production spectrum, also determined via pQCD, and vacuum fragmentation functions.  While many pQCD-based energy loss calculations describe the suppression of neutral mesons at RHIC, $R_{AA}^{\pi^0}$, none are able to simultaneously describe quantitatively both the normalization and angular distribution of $R_{AA}$ of $\pi^0$'s nor the normalization of $R_{AA}$ for both $\pi^0$'s and heavy flavor electrons.  We can quantify the generic statement about over-suppression given above using the WHDG energy loss model \cite{Wicks:2005gt}, which includes both the collisional and radiative energy loss processes in a realistic, Bjorken-expanding medium: predictions for the suppression of high-$p_T$ $\pi^0$ mesons assuming that the QGP density scales with the observed charged particle multiplicity, $dN_{ch}/d\eta$, with the proportionality constant between the QGP density and $dN_{ch}/d\eta$ found by a rigorous statistical analysis \cite{Adare:2008cg}, are given in \cite{Horowitz:2011gd}; these predictions are compared to the published ALICE data \cite{Aamodt:2010jd} in Fig.~\ref{RAA} (a) and preliminary CMS data \cite{CMSRAA} in Fig.~\ref{RAA} (b).  One sees in Fig.~\ref{RAA} (a) that despite the over factor of 2 increase in medium density, the observed increase in suppression in published ALICE data from top RHIC energy to LHC is small; there is even only a small increase in suppression from preliminary $\surd s$ = 62.4 AGeV data \cite{Purschke} to $\surd s$ = 200 AGeV data at RHIC.  Preliminary CMS data, with $R_{AA}$ out to $p_T\sim100$ GeV/c, shows a significant rise as a function of $p_T$, which is in exact qualitative agreement with pQCD expectations: perturbatively, the fractional energy loss of high-$p_T$ particles goes as $\epsilon = (E_f-E_i)/E_i \sim \log(p_T)/p_T$ and $R_{AA}\sim(1-\epsilon)^n$, where $n+1\sim5$ is the power that best approximates the power law production spectrum for partons at LHC.  Quantitatively, the parameter-free WHDG prediction at LHC energies rigorously constrained by RHIC data is significantly more suppressed for 0-5\% centrality collisions than the central values of the published ALICE data.  As the WHDG predictions are consistent with the peripheral 70-80\% collisions, the WHDG $R_{cp}$ does not compare well with the that which one extracts from the ALICE data, especially as the uncertainty in the production spectra drops out of the experimental results.  However, the comparison between the WHDG predictions and the preliminary CMS data in Fig.~\ref{RAA} (b) is quite reasonable with the given experimental and theoretical uncertainties.  It seems, then, that the quantitative question of whether the WHDG calculation, and pQCD-based energy loss calculations in general, can describe the normalization of $R_{AA}$ at LHC when constrained by RHIC is still open.  Hopefully the answer to this question will become clearer when the p+p baseline spectrum is analyzed out to $p_T\sim100$ GeV/c.  However, there are two additional questions: 1) to what extent do initial state effects, such as those that turn off as a function of $p_T$ (due to, e.g., the suppression of small-$x$ gluons as expected from saturation physics), account for the qualitative rise in $R_{AA}(p_T)$ at LHC? and 2) can WHDG, and pQCD-based energy loss calculations in general, describe quantitatively describe multiple observables (e.g. $v_2$, the suppression of heavy flavors, $I_{AA}$, etc.) out to very large, $p_T\sim100$ GeV/c?  We look forward to the answer of the first question, which will come from the observation of $p_T\gtrsim5$ GeV/c direct photons and/or a measurement of suppression in p + A collisions.  The second question has been preliminarily addressed at moderate momenta in Fig.~\ref{v2RAA}: the parameter-free WHDG extrapolation to LHC provides an excellent description of (a) $v_2(p_T)$ of charged hadrons measured by ATLAS \cite{Jia} and (b) the suppression of $D$ mesons as measured by ALICE \cite{Dainese:2011vb}.  Note that the suppression of $D$ mesons begins to exceed that of pions at $p_T\sim20$ GeV/c, due to the much more steeply falling production spectrum and the shortened formation time of the heavy quark.  It will be interesting to see if the agreement seen in Fig.~\ref{v2RAA} continues once the uncertainties and momentum reach of the measurements improve.  The preliminary results from CMS on the distribution of the energy lost from a high-$p_T$ particle to very wide angles also helps constrain the energy loss mechanism in QGP.  pQCD intuition would suggest that perturbative radiative energy loss would be concentrated at collinear angles of $\sim\mu/E$, a more detailed examination of the differential single inclusive gluon radiation distribution shows that pQCD predicts the emission of radiation at quite large angles \cite{Horowitz:2009eb}.

\begin{figure}[!htb]
\centering
$\begin{array}{cc}
\includegraphics[width=2in]{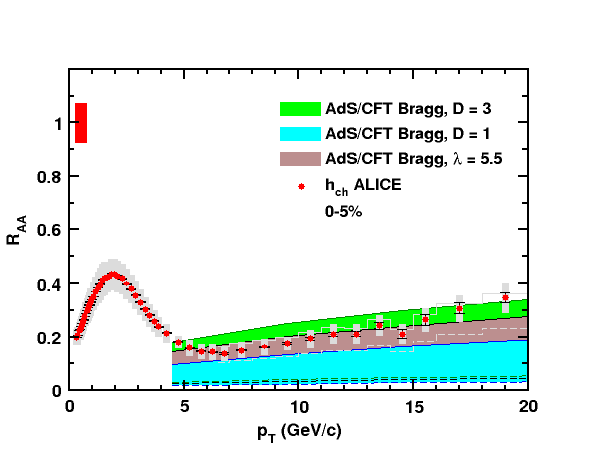} &
\includegraphics[width=2.5in]{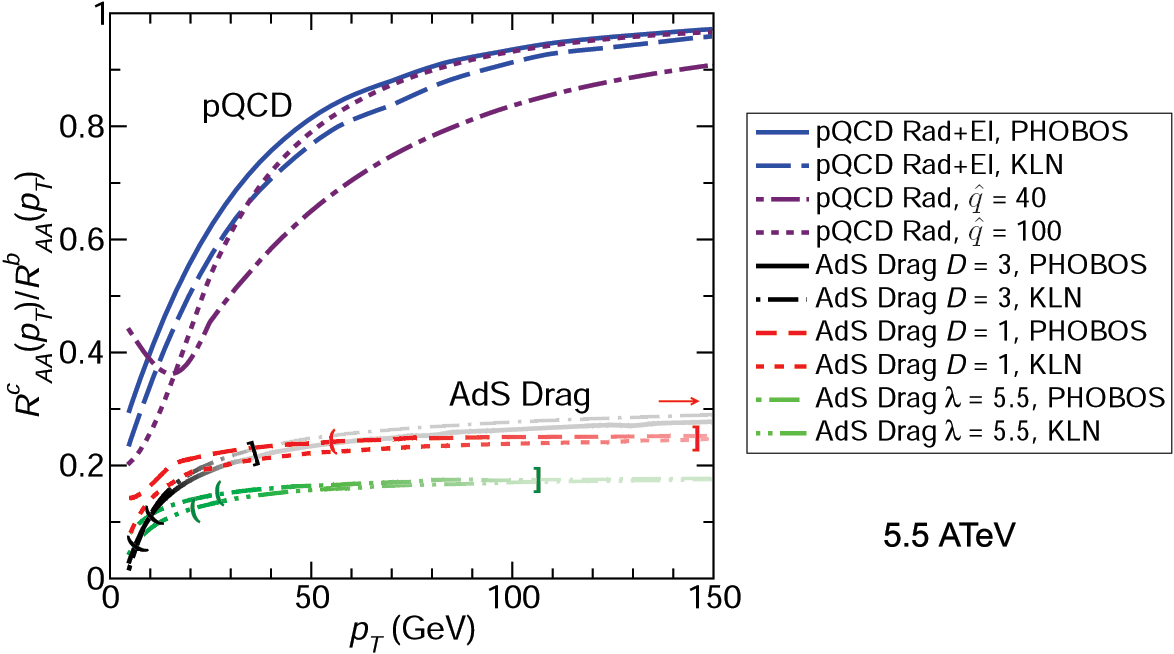}
\end{array}$\\
\begin{flushleft}
\vspace{-.2in}
$\begin{array}{ll}
\mbox{\hspace{.5in}\footnotesize(a)} & \mbox{\hspace{2.2in}\footnotesize(b)}
\end{array}$
\end{flushleft}
\vspace{-.1in}
\caption{\label{AdS}(a) Comparison of the AdS/CFT Bragg model to ALICE charged hadron $R_{AA}(p_T)$ data. (b) Predictions of the double ratio of $c$ to $b$ quark $R_{AA}(p_T)$ at 5.5 ATeV from fully 1) weakly-coupled and 2) strongly-coupled energy loss models.
}
\end{figure}

\emph{Nonperturbative Energy Loss and Comparison} Fully nonperturbative treatments of heavy quark energy loss at RHIC show qualitative agreement with the measured suppression of non-photonic electrons \cite{Horowitz:2010dm}.  We show in Fig.~\ref{AdS} (a) that a simple Bragg model of light quark and gluon energy loss, in which the probability of escape for a parent parton is given by $P_{escape}(L) = \theta(L_{therm}-L)$, where $L_{therm} \sim E^{1/3}$ (with appropriate proportionality coefficients found in \cite{Chesler:2008uy}) provides a qualitatively consistent picture of the measured suppression of charged hadrons at LHC.  It is important to note that the pQCD energy loss calculations appear to have a strong dependence on the initial thermalization time and pre-thermalization conditions \cite{Buzzatti:2011vt}.  Additionally, although there may be future means of directly measuring the initial gluon wavefunction at an electron-ion collider \cite{Horowitz:2011jx}, model calculations of observables are also dependent on the medium geometry used.  We therefore suggest that the double ratio of charm to bottom quark $R_{AA}(p_T)$ will provide valuable---perhaps even unambiguous---evidence of the dominant energy loss mechanism at work in heavy ion collisions, with qualitatively different behavior for fully weakly and fully strongly coupled energy loss calculations.  In particular, perturbative QCD predicts a rapid rise to 1 in the ratio as a function of $p_T$ as the pQCD calculation becomes insensitive to the mass of the parent parton; on the other hand, AdS/CFT drag results suggest a nearly $p_T$-independent ratio significantly suppressed from 1 at approximately the ratio of the quark masses, $m_c/m_b$.  

\end{document}